\documentclass[aps,prl,reprint,superscriptaddress,amsmath,amssymb,floatfix]{revtex4-2}
\usepackage{graphicx}
\usepackage[dvipsnames]{xcolor}
\usepackage[utf8]{inputenc}
\usepackage[T1]{fontenc}
\usepackage{hyperref}

\begin{document}

\title{Functional understanding of structural components in polypseudorotaxane gels}

\author{Ella Sapsford}
\affiliation{School of Physics and Astronomy, University of Edinburgh, Edinburgh, EH9 3FD, UK}
\affiliation{School of Chemistry, University of Edinburgh, Edinburgh, EH9 3FJ, UK}
\author{Andrea Bonato}
\affiliation{School of Physics and Astronomy, University of Edinburgh, Edinburgh, EH9 3FD, UK}
\author{Simon Weir}
\affiliation{School of Physics and Astronomy, University of Edinburgh, Edinburgh, EH9 3FD, UK}
\author{Nawar Al-Zebari}
\affiliation{Reckitt Benckiser Health Limited, 103--105 Bath Road, Slough, Berkshire, SL1 3UH, UK}
\author{Paul J. Lusby}
\affiliation{School of Chemistry, University of Edinburgh, Edinburgh, EH9 3FJ, UK}
\author{Chenfeng Ke}
\affiliation{Department of Chemistry, Washington University in St.\ Louis, One Brookings Drive, Saint Louis, Missouri 63130, USA}
\author{Davide Michieletto}
\email{Davide.Michieletto@ed.ac.uk}
\affiliation{School of Physics and Astronomy, University of Edinburgh, Edinburgh, EH9 3FD, UK}
\affiliation{Institute of Genetics and Cancer, University of Edinburgh, Edinburgh, EH4 1QY, UK}
\affiliation{International Institute for Sustainability with Knotted Chiral Meta Matter (WPI-SKCM$^2$), Hiroshima University, Higashi-Hiroshima, Hiroshima 739-8526, Japan}

\date{\today}

\begin{abstract}
Biomedical devices, wearable interfaces, and soft robots require compliant soft materials with reliable mechanical performance, yet the molecular principles needed to design such materials are not always fully understood. Thus, there is an urgent need to understand how to design novel stretchable and soft hydrogels with enhanced toughness, i.e. resistance to breakage. Slide-ring gels (SRGs) are a class of unconventional soft materials formed by polyrotaxanes (PR) building blocks whereby sliding cyclodextrins (CD) replace traditional chemical and physical crosslinks and in turn enhance toughness and stretchability by several orders of magnitude. The rational design of SRGs remains challenging due to a lack of a mechanistic understanding of the role played by individual components. 
In this work, we implement simple synthetic perturbations that individually target polymer-CD and CD-CD interactions in unstoppered pseudopolyrotaxanes (pPR) solutions and measure the consequent changes in viscoelastic properties. Weakening the intermolecular interaction between CDs abolishes crystallisation and significantly weakens the gel. At the same time, thermodynamically favouring threading by using azobenzene-modified PEG chains favours the formation of stable and long-lived networks even at remarkably low CD stoichiometries. We rationalise these observations with coarse-grained Molecular Dynamics simulations which clarify the structure-function relationship of the components forming these materials.
Our work contributes to guiding the rational design and development of enhanced SRG hydrogels and elastomers.
\end{abstract}

\keywords{slide-ring gel, structure-function, rheology}

\maketitle

\begin{figure*}[t]
    \centering
    \includegraphics[width=0.75\linewidth]{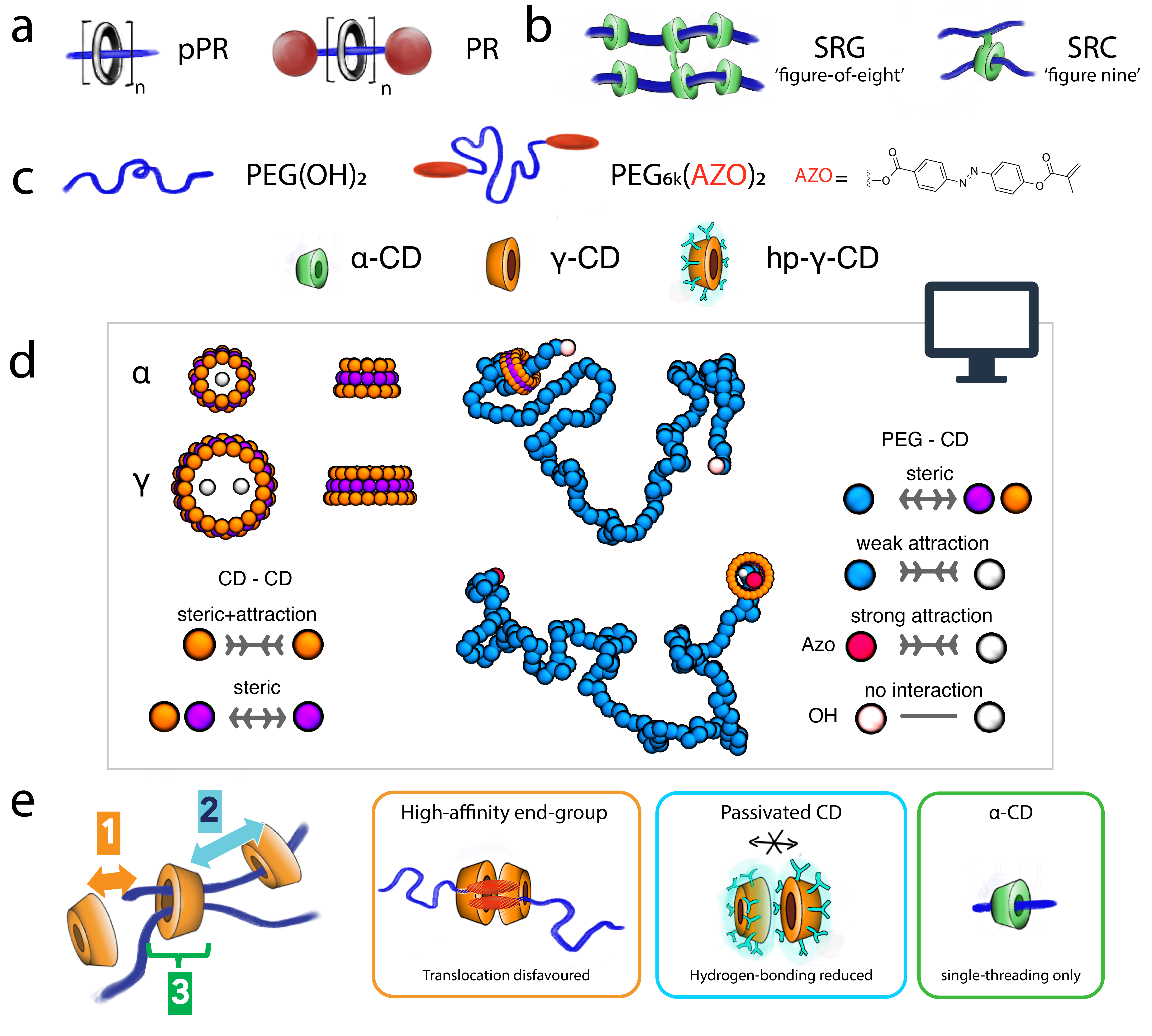}
    \caption{\textbf{Components of pseudopolyrotaxanes (pPRs) in simulations and experiments.} \textbf{(a)} Schematic definitions distinguishing pseudopolyrotaxanes (pPRs) from polyrotaxanes (PRs). \textbf{(b)} Schematic definition of 'figure of eight' crosslinks in slide-ring gels (SRG) and 'figure nine' crosslinks found in some slide-ring crosslink (SRC) gels. \textbf{(c)} Schematic illustration of the synthetic modifications made to each structural component. \textbf{(d)} Representative components of the coarse-grained bead-spring simulation; interactions between bead types are defined by colours. \textbf{(e)} Single component modifications within the pPR system: naitve vs functionalised end groups, passivated vs interacting CDs and $alpha$ vs $\gamma$ CDs.}
    \label{fig:1}
\end{figure*}

\section{Introduction}
Common hydrogels, i.e. gels swollen in water, are typcally brittle and inextensible, features that strongly limit their practical applications~\cite{Li2024}. Due to their physiological compatibility and widespread use, there is an acute need to produce so-called ``tough'' hydrogels that display reduced embrittlement while retaining softness and low solid content~\cite{Gong2003, DasMahapatra2020, Sun2012}. 
One solution to this challenge is the use of poly(rotaxanes) (PR) components made of small ring molecules, often cyclodextrins (CD), that are threaded by polymer chains, typically poly(ethylene glycol) (PEG)~\cite{Harada1992, Wenz2006} 
(see Fig.~\ref{fig:1}). Slide-ring gels (SRG) are derivatives of PR, and are formed by covalently crosslinking CDs in the system. SRGs display remarkable extensibility and toughness due to the CDs mobility along the chains: upon applied strain, they act as mobile crosslinks and can readily redistribute the external applied stress~\cite{Ito2007}. More specifically, SRGs with low coverage of CDs -- i.e. a small number of CDs per chain -- are markedly tougher than their high coverage equivalents thanks to the well-known ``pulley mechanism''~\cite{Karino2006,Ito2007}.
PRs can also be covalently bonded within more traditional hydrogel copolymer networks to improve their mechanical properties and in turn form slide-ring crosslinker (SRC) materials~\cite{BinImran2014, Zheng2023, Koyanagi2017}.
For example, SRCs can form gels through the covalent linking of CD sliding rings to a hydrogel copolymer~\cite{BinImran2014}. These ``figure nine'' crosslinks, Fig.~\ref{fig:1}b, have demonstrated mixed mechanical performance, and are usually vastly inferior to elastomeric SRG examples consisting of no threading hydrogel copolymer. 
In contrast to this, the work of Ke et al.~\cite{Tang2023}, utilizes gamma-cyclodextrin ($\gamma$-CD) in turn achieving a  ``figure zero'' crosslinking motif that can thread two chains within its larger cavity~\cite{Harada1994}. The use of $\gamma$-CD is expected to simplify the synthesis of SRG. Indeed, because $\gamma$-CD can accommodate two chains, it is not necessary to perform further chemical conjugation to form a crosslink
~\cite{Harada1994,Tang2023,Zheng2022,Sapsford2025}.

The double threading of $\gamma$-CD has
has been known since the 1990s~\cite{Harada1994}. However, applications of the ``figure zero'' crosslinking motif in SRGs have thus far been limited~\cite{Sapsford2025}. The mechanical and rheological implications of using $\gamma$-CD to crosslink SRGs have not been thoroughly studied and understood. 
Additionally, the majority of current methods to form SRGs rely on the formation of a pseudo-poly(rotaxane) (pPR) precipitate formed through the (slow) crystallisation of CD domains during the (fast) chain-threading process~\cite{Fleury2005}. The slow crystallisation impacts the threaded coverage, resulting in a poorly controllable degree of threading, typically in the range between 10-25\%~\cite{Lin2021,Ito2007}. This has a detrimental effect on the ``pulley mechanism'' which requires low coverage 
($\lesssim$5\%) to work optimally~\cite{Jiang2018,Kato2015}.

To achieve better control over the synthesis of SRCs with desired mechanical and rheological properties, here we dissect the role of different components of the gels. 
We combine both experimental chemical design to independently tune the network's components and coarse grained simulations to elucidate their roles on the resulting material. Analysis of these simulations allow us to gain fine structural insights into the internal arrangement of the networks. Our combination of experimental and \emph{in silico} work creates a unique correspondence between mechanical and (micro)structural properties of pPR gels. 

\section{Results and Discussion}

PR are mechanically ``interlocked'' units -- similar to microscopic beaded necklaces~\cite{Harada1992} -- which can enhance stretchability and mechanical properties of materials through microscopic ``pulley'' mechanism~\cite{Ito2007}. Here, however, we focus on pseudo-PR (pPR) where the polymeric components are not stoppered. This choice of focus is two-fold: Firstly, SRCs are poorly defined, intrinsically heterogenous, and undergo a uncontrolled polymerisation of the copolymer yielding a complex network structure. Secondly, the implementation of a sufficiently bulky, water-soluble stopper is not trivial. 
Therefore, for simplicity and reproducibility, we focus on pPRs -- which already display a rich range of incompletely understood rheological behaviours -- with the aim of extending these results to SRG/SRCs. 
 
Within pPRs, intermolecular interactions are known to be influential on the threading of PEG chains through the CDs~\cite{Cai2023, Okumura2000}. We study these interactions by enhancing the thermodynamic stability of the chain-end with the CD, by abolishing the CD-CD interaction and by examining the difference between using $\gamma$ (potentially double threading) or $alpha$ (single-threading) CD in the pPR system. 
These ``single component perturbations'' were realised as follows: native PEG(OH)$_2$ was replaced by telechelic azobenzene-PEG (PEG(\textbf{Azo})$_2$), native $\gamma$-CD with its ``passivated'' counterpart hydroxy-propyl $\gamma$-CD (hp-$\gamma$-CD), and finally $\alpha$-CD replaced $\gamma$-CD (see Fig.~\ref{fig:1}b, d).

We employed bulk rheology of the resulting complex fluids to understand the impact of these components on the viscoelasticity of the fluid and  molecular dynamics (MD) simulations of PEGs and cyclodextrins in bulk to quantify their internal microstructure. PEGs were modelled as bead-spring polymers, whereas CDs as rigid complexes composed of three stacked, concentric rings of beads (Fig.~\ref{fig:1}c). To simulate CD crystallisation, we introduced attractive interactions between the beads belonging to the outer rims (the two outer rings) of different CDs; for passivated CDs, this attraction is turned off. To capture the hydrophobic effect between PEGs and the inner cavity of the CDs, we placed either one (for $\alpha$-CDs) or two (for $\gamma$-CDs) ``phantom'' beads at the centre of each CD (grey, Fig.~\ref{fig:1}c). We then set a weak attraction between the PEG beads and these phantom beads and a strong attraction these beads and the Azo groups, set as the terminal beads of the PEG(Azo)$_2$ (see Fig.~\ref{fig:1} and SI for more details). Finally, MD simulations informed microstructural details and elucidated causation for the observed rheological behaviour.

\subsection{Thermodynamically enhanced threading improves elasticity}

\begin{figure*}[t]
    \centering
\includegraphics[width=0.8\linewidth]{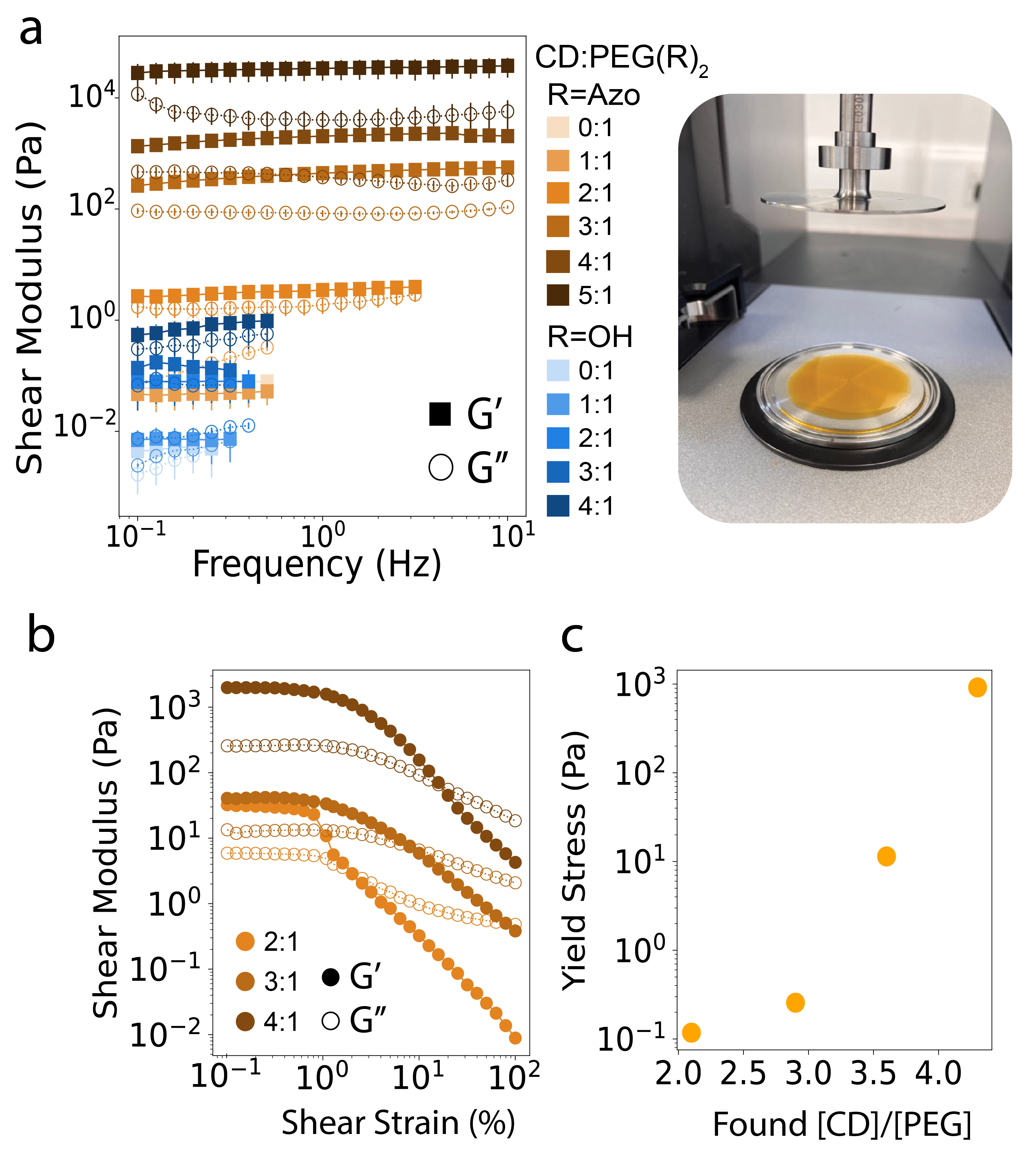}
    \caption{\textbf{Rheology of pseudopolyrotaxane (pPR) gels with thermodynamically enhanced $\gamma$-cyclodextrin ($\gamma$-CD) threading.} \textbf{(a)} Oscillatory frequency sweeps (0.1–10 Hz, 0.2\% strain) of pPR gels formed from PEG(OH)$_2$ (blue) and PEG(Azo)$_2$ (orange) across a range of low-threading compositions; image shows the loaded pre-gel PEG(Azo)$_2$ sample before rheological testing. \textbf{(b)} Oscillatory amplitude sweeps (1 Hz, 0.1–100\% strain) for PEG(Azo)$_2$ pPRs at CD:PEG ratios of 2:1 to 4:1. \textbf{(c)} Yield stress, taken as the shear stress at which $G^\prime$ reaches 90\% of its plateau value (see Fig.~S10), plotted as a function of CD:PEG ratio- original fed ratio is 2:1-5:1 ([CD]/[PEG]).}
    \label{fig:2}
\end{figure*}

\begin{figure*}[t]
    \centering
    \includegraphics[width=1\linewidth]{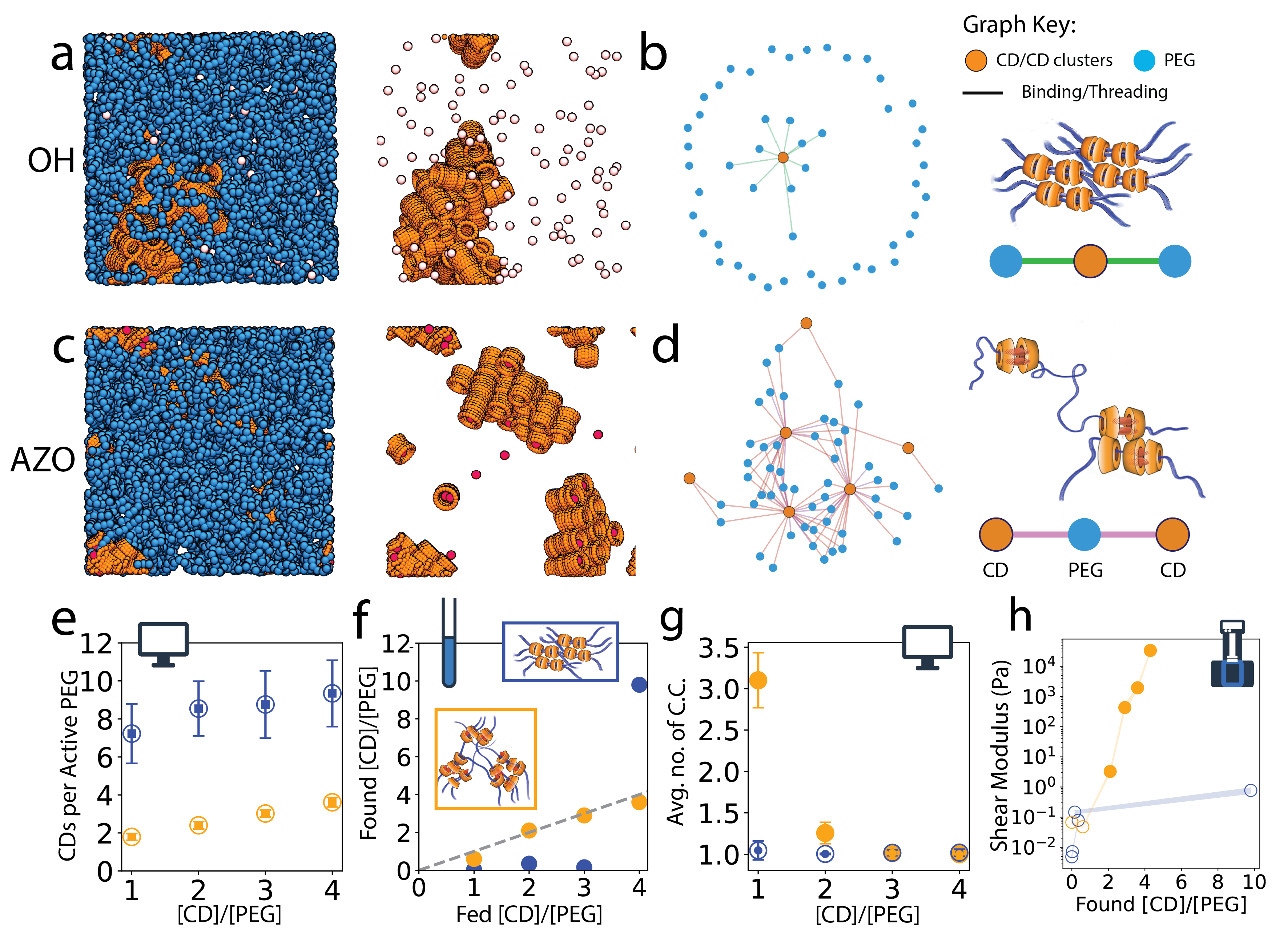}
    \caption{\textbf{Thermodynamically enhanced PEG(Azo)$_2$ yields microphase separation in $\gamma$-cyclodextrin ($\gamma$-CD) pPR gels.} \textbf{(a)} Snapshots from MD simulations of $\gamma$-CDs (orange) and PEG(OH)$_2$ (blue and white; OH groups in white).  CD:PEG ratio is 3. \textbf{(b)} Graph representation of the configuration in \textbf{a}. CD clusters are represented as orange nodes and PEGs as blue nodes. An edge connects a PEG node to a CD cluster if the PEG threads through a CD belonging to that cluster. Schematic depiction of aggregative building block. \textbf{(c-d)} Same as \textbf{(a-b)}, but for PEG(\textbf{Azo})$_2$. Red beads model the terminal Azo groups. In the graph in \textbf{(d)}, edges connect a PEG to a CD cluster if at least one of its two Azo terminals threads/binds to any element in the cluster. 
    Schematic depiction of bridging (macro)monomer. \textbf{(e)} Number of threaded CDs per PEG, only ``active'' PEGs that are either threading or binding are included in the statistics. \textbf{(f)} Ratio of fed CD against ratio of found CD that have become threaded measured by $^1$H NMR. The dashed line in the plot represents Found = Fed relationship, which would be achieved for perfect incorporation of CDs into the pPR network. \textbf{(g)} Degree of percolation of network, defined as the average number of connected components in the simulated network at steady state, as a function of different [CD] to [PEG] ratio. \textbf{(h)} Stiffness as a function of the found [CD]:[PEG] ratio, measured via  $^1$H NMR.}
    \label{fig:3}
\end{figure*}

To understand the role played by the interactions between chain-end and $\gamma$-CD, we compared pPR systems in which the chain end is functionalised with an azobenzene derivative. This functional group is known to bind $\gamma$-CD more strongly than $\alpha$-CD~\cite{Tang2023}. We prepare pPR solutions by mixing $\gamma$-CD with either PEG(OH)$_2$ or PEG(\textbf{Azo})$_2$ and measure their mechanics by small-angle oscillatory rheology (frequency and amplitude sweeps). 

pPRs obtained from PEG(OH)$_2$ display only a very narrow window in which a plateau modulus may be determined. As the equivalents of $\gamma$-CD are increased the stiffness of the elastic fluid increases, however the pPR fail to form a gel at these low feed ratios. The fluids also phase separate at higher frequencies, and even without perturbations at long times after the initial mixing. This bulk phase separation is not compatible with rheological characterisation (see Fig.~S8 in SI).  

In contrast, pPR made with a mixture of PEG(\textbf{Azo})$_2$ and $\gamma$-CD display a remarkable elastic behaviour even at low feed ratio of 2:1 (see fig.~\ref{fig:2}b) and a broad range of elastically-dominated frequencies where $G^\prime > G^{\prime \prime}$. The plateau modulus of the resulting gel state increases by nearly an order of magnitude for each additional equivalence of $\gamma$-CD (Fig.\ref{fig:2}a). 

To understand the ability of pPR gels to resist deformations, we performed  amplitude sweeps in the linear viscoelastic regime (VLE), i.e. at fixed frequency of 1 Hz, Fig.~\ref{fig:2}b-c. The yield stress, i.e. the stress measured on the rheometer at which the gel-like state ``yields'' 
defined as 90\% of the plateau modulus (Fig.~S10), indicates a marked change in behaviour between 3:1 and 4:1 molar ratios. Above three $\gamma$-CDs per PEG chain, the yield stress increases exponentially suggesting an Arrhenius-like behaviour. The yield stress can be physically interpreted as the stress at which the structural crosslinks within the gel are ``broken'', allowing the PEG chains to slide and the solution to flow (Fig.~\ref{fig:2}g). 

The propensity for gel formation with PEG(\textbf{Azo})$_2$ suggests a structural difference in the pPR when compared to PEG(OH)$_2$. 
The high-binding Azo affinity group favours $\gamma$-CD threading at the chain ends. The importance of the location as well as the number of threaded rings within PR gels has been demonstrated in previous work on both kinetically-trapped~\cite{Lin2021} and block copolymer PR systems~\cite{Uenuma2019}.

To better understand the structural difference we performed MD simulations. Fig.~\ref{fig:3}a shows that pPRs constructed with no additional interaction at the chain-end, i.e in the PEG(OH)$_2$ case, are bulk separated. The network connectivity in these simulations can be displayed by graph representation (Fig.\ref{fig:3}b). Specifically, orange points are CD clusters and blue points represent each PEG chain. Fig.~\ref{fig:3}b shows one large pPR cluster that is threaded by many PEG chains. This ``aggregative'' morphology can be thought of as a PEG-CD-PEG representation: the system forms only one large ``sticky'' region and cannot display elasticity. Instead, the pPRs display macrophase separation and liquid-like behaviour is expected.

To model the stronger binding of PEG(Azo)$_2$ to $\gamma$-CD, we simulated polymers whose ends are strongly attracted to the centre of the coarse-grained CD (respective to the blue PEG bead-CD centre attraction). In this case the simulations display a micro-phase separated state made of small CD clusters connected by PEG chains (see Fig.~\ref{fig:3}c). This phase is characterised by CDs arranging in honeycomb crystalline regions, which agrees well with experimental scattering data of similar pPRs~\cite{Tang2023}. This agreement suggests that our relatively simple coarse grained model captures the complex behaviour of CD-PEG pPR systems. 

As before, we can characterise the microscopic structure formed by pPRs by using a network representation: In Fig.~\ref{fig:3}d, connections between $\gamma$-CD clusters and PEG chains can be drawn if a PEG is bound to, or threading through, a CD. At a low CD-to-PEG ratio ([CD]:[PEG]=3), we observe that all nodes in the network are connected, i.e., the system is fully percolated and therefore can support mechanical stress. 

Through our simulations, we can also count the number of CDs that are threaded or bound by PEGs and compare it with experiments. PEGs that are involved in threading or binding CDs are defined as ``active'' in what follows. The number of CDs per active PEG found by simulations strongly depends on the strength of the interaction between the chain-end and CDs (Fig.~\ref{fig:3}e). In the PEG(OH)$_2$ case only some of the PEGs are active, and most of the CDs are threaded by few PEG chains (Fig.~\ref{fig:3}b and SI, Fig.~S17d); however, in the PEG(Azo)$_2$ case, we observe a more even distribution of CDs per PEG chain, which contributes to the network becoming percolated.

To compare with the simulations, in Fig.~\ref{fig:3}f we show the experimentally measured [$\gamma$-CD]:[PEG] ``found'' threading ratio, determined by $^1$H NMR on the precipitated pPR. Fig.~\ref{fig:3}f has two features worthy of note: firstly, in the PEG(\textbf{Azo})$_2$ case, simulations show a remarkable agreement with experimental values, where this found ratio also scales linearly with the feed ratio; secondly, in the PEG(OH)$_2$ case, the obtained '`found'' ratios are much lower than the CDs per active chain as determined by simulations (Fig.~\ref{fig:3}e). 
The variation between the experimental and \emph{in silico} data in PEG(OH)$_2$ case could also be due to dethreading events during the extraction of the pPRs. Dethreading is especially likely for singly threaded $\gamma$-CD as the binding affinity is very low. Additionally, the PEG(OH)$_2$ system displays an abrupt jump at fed [CD]/[PEG]=4 which suggests strong cooperativity, i.e. at this feed ratio, threading of $\gamma$-CDs onto a PEG chain leads to more threading onto the same chain; this cooperativity in threading was predicted in Ref.~\cite{Okumura2000} as due to the polymeric substrate ``templating'' and facilitating the crystallisation of CD clusters.

Further computational analysis of the network structure in the simulations show that in the high-binding affinity case, percolation is enabled at  3:1 [CD]:[PEG] (Fig.\ref{fig:3}g, SI~16c), meaning that the network is gel forming. Conversely, where unfunctionalised, pPRs 
display macro-phase separation 
across all tested ratios.

Importantly, the results in Fig.~\ref{fig:3}e-f also strongly suggest that end-functionalisation of PEG polymers with a high binding affinity group (such as Azo) is a robust method to control the number of CDs threading PEG chains.

In summary, in this section we discovered that including a high-affinity Azo group at the ends of PEG chains contributes to the stable formation of CD-PEG-CD (macro)monomers capable of efficient network formation. Similar results have been seen in previous works with kinetic, rather than thermodynamic, trapping of $\gamma$-CD~\cite{Liu2021, Smith2024}. 
The pinned localisation of $\gamma$-CDs at the end of the PEG chains turns the chains into A-B-A-like triblock co-polymers, with the A block being multivalent sticky regions formed by $\gamma$-CD clusters connected by PEG chains. 
The increased stability of $\gamma$-CDs at the chain-end also means that translocation of $\gamma$-CDs is restricted, in turn arresting macro-phase separation and precipitation. 
Arrested, or micro-phase, separation leads to the formation of a space-filling network that is kinetically trapped even in the absence of stoppers. 

Key for the design of efficient SRGs, we demonstrated that the azobenzene functional group allows the formation of pPR gels at extremely low CD:PEG ratios, thus indicating a potential strategy for the assembly of highly stretchable low coverage SRGs.

\subsection{Passivated $\gamma$-CDs abolish gelation of pPRs even in the presence of Azo groups}

\begin{figure*}[t]
    \centering
    \includegraphics[width=0.9\linewidth]{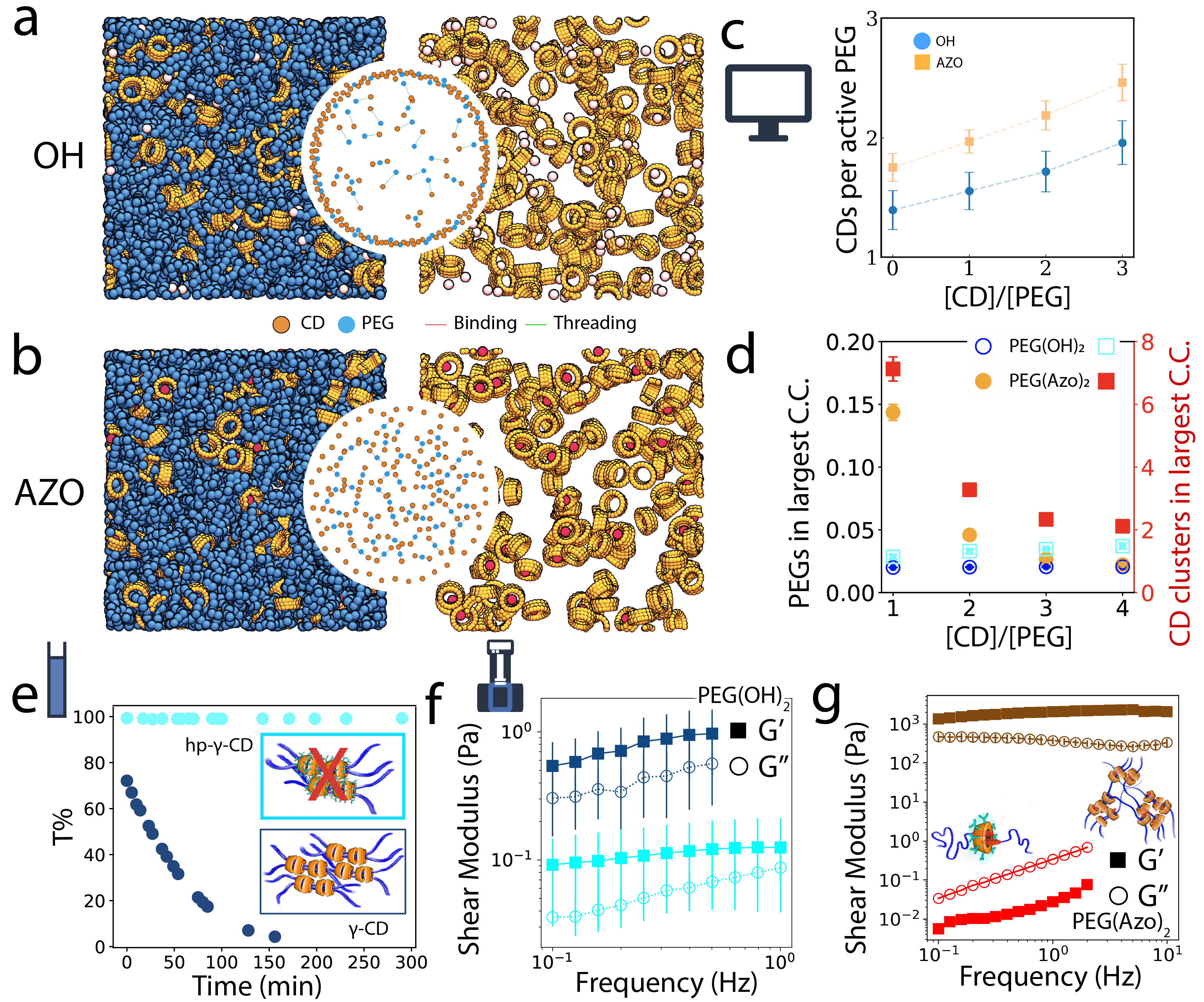}
    \caption{\textbf{Passivated hp-$\gamma$-CD abolish gelation.} \textbf{(a)} Snapshots and graph depiction from simulations with hp-$\gamma$-CD (yellow) with PEG(OH)$_2$. Graph key is the same as Fig.~\ref{fig:3}. \textbf{(b)} Same as \textbf{a} for PEG(\textbf{Azo})$_2$. \textbf{(c)} Number of threaded CDs per PEG, only ``active'' PEGs that are either threading or binding are included in the statistics. \textbf{(d)} Average fraction of PEG and number of CD clusters in the largest connected component (C.C) -- PEGs (left black axis, circle points), CD clusters (right red axis, square points) for PEG(OH)$_2$ (blue, cyan) and PEG(Azo)$_2$ (orange, red). \textbf{(e)} Turbidity experiment monitoring the light transmittance at wavelength $\lambda$ = 550 nm quantifies the formation of large aggregates in mixtures of PEG(OH)$_2$ and $\gamma$-CD. Absence of turbidity indicates that mixtures of PEG(OH)$_2$ with passivated hp-$\gamma$-CD  do not form aggregates (\textit{cyan}).  \textbf{(f)} Rheological frequency sweep for native PEG(OH)$_2$ with native (navy) and passivated hydroxypropyl (hp-) (cyan) $\gamma$-CDs. \textbf{(g)} Rheological frequency sweep for PEG(Azo)$_2$ with native (brown) and passivated hp- (red) $\gamma$-CD.}
    \label{fig:4}
\end{figure*}

In the previous section, we observed that clustering of $\gamma$-CD contribute to precipitation of the PEG(OH)$_2$ system and gelation of the PEG(Azo)$_2$ system. Additionally, we know from previous work that CD-CD attraction affects the threading process~\cite{Chen2023, Cai2023, Noritomi2022}. Indeed, when a ring becomes threaded, hydrogen bonds between the exposed hydroxyl groups on the CDs ``faces'' aid the recruitment of additional rings on the same polymer backbone~\cite{Fleury2005}. The threaded polymer then templates the CDs, which in turn nucleate the formation of crystal domains between or along the polymer(s). Where these crystalline  regions are between chains they may be crosslinking.  Notably, this is the only mode for $\alpha$-CD to create a physically crosslinked network without additional chemical steps to covalently bond CDs.

When the CD-CD attraction is weakened, the threading interaction loses cooperativity~\cite{Okumura2000}, the found threading ratio becomes proportional to the feed ratio~\cite{Noritomi2022}, and crystallisation is disfavoured. Weakening the inter-CD interaction is therefore expected to affect the rheology of the pPR fluids, and can be achieved by chemical modification of the CD's exposed hydroxyl groups, such as through hydroxypropyl (hp-) modifications (Fig.~\ref{fig:1}). The hp- group disrupts hydrogen bonding and we thus dub hp-$\gamma$-CDs as ``passivated'' $\gamma$-CDs (Figure \ref{fig:3}a). In this section we therefore investigate how passivating CDs affect the rheology and behaviour of pPR complex fluids. 

To simulate the passivated $\gamma$-CDs we turn off the attractive part of the interaction between the beads forming the rings (see SI.~Simulations Set up for more details). As shown in Fig.~\ref{fig:4}, no crystallisation of the $\gamma$-CDs is observed in either the PEG(\textbf{Azo})$_2$ and PEG(OH)$_2$ cases. Instead, we observe that a small number of passivated $\gamma$-CD become threaded and that the system displays no cooperativity (see SI. Fig.~S19a-b). At low [CD]:[PEG] ratios the increased binding affinity acts to recruit slightly more rings to the PEG chains (Fig.~\ref{fig:4}c), however, the thermodynamically enhanced interaction is not strong enough to promote supramolecular polymerisation over a range of CD ratios (see SI Fig.~S19). Additional equivalents of passive $\gamma$-CD then compete with each other to occupy the chains, and by 4:1 [CD]:[PEG] ratios, there is only $\sim 1-2$ chains per cluster (Fig.~\ref{fig:4}b, d).
Therefore, without the sticky interactions present in native $\gamma$-CD we expect our systems will no longer be gel-forming.

To experimentally test our simulated predictions we first check if hp-$\gamma$-CDs form crystalline domains. To do this, we performed turbidity experiments, in which the transmitted light at wavelength 550nm through pPR mixtures of PEG and either native or hp-$\gamma$-CD was measured (Fig.~\ref{fig:4}e). We observe that the samples with passivated $\gamma$-CDs do not display the decrease in transmittance associated with CD clustering in native $\gamma$-CD samples, even after twice the time-period, thus confirming the expected lack of CD clustering (Fig.~\ref{fig:4}d).  

According to the simulations, we also expect that the lack of crystalline domains reduces physical crosslinking between PEG chains. This hypothesis was tested through rheological measurements for the PEG(OH)$_2$ (Fig.~\ref{fig:4}e) and PEG(Azo)$_2$ (Fig.~\ref{fig:4}f).  For fixed CD:PEG(OH)$_2$ feed ratio, the rheology measurements indicate that passivating CDs yields a significant loss in stiffness, and $G^\prime \simeq G^{\prime \prime} \simeq 0.05-0.1$ Pa (Fig.~\ref{fig:4}e). The loss of elasticity and viscosity is even more pronounced in the PEG(\textbf{Azo})$_2$ systems, with a drop of 5 orders of magnitude for passivated $\gamma$-CDs (Fig.~\ref{fig:3}f). Additionally, the viscous modulus is about 10-fold larger than the elastic modulus across frequencies, indicating a clear liquid-like behaviour and no observable elastic-dominated regime in the presence of hp-$\gamma$-CD. 

In this section, we thus discovered that pPR formed with passivated $\gamma$-CD cannot form gels even in the presence of high-binding end groups. This evidence follows our hypothesis that the crosslinking in pPR gels is overwhelmingly derived from the formation of crystalline regions through short-ranged CD-CD interactions. 

\subsection{pPRs made with $\alpha$-CDs display gel forming properties at higher equivalences}

\begin{figure*}[t]
    \centering
    \includegraphics[width=0.8\linewidth]{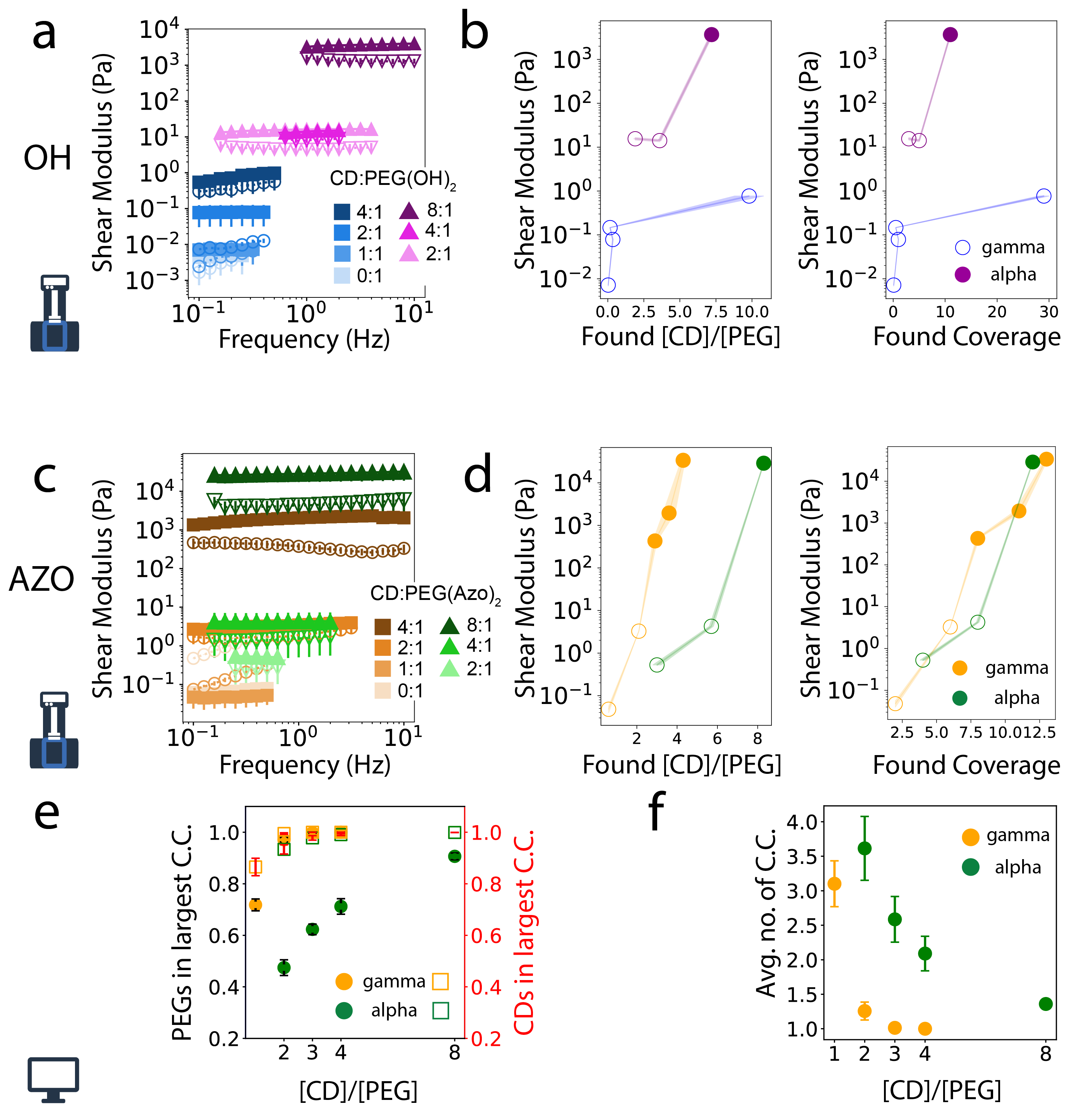}
    \caption{\textbf{Network structure is insensitive to the cyclodextrin (CD) size.} \textbf{(a)} Rheological frequency sweep for native PEG(OH)$_2$ with $\gamma$-CD (blue) and $\alpha$-CD (purple)(0.1–10 Hz, 0.2\% strain). \textbf{(b)} Stiffness as a function of the found [CD]:[PEG] ratio (left) and coverage (right), measured via  $^1$H NMR. Coverage refers to the percentage of the polymer chain that is ``covered'', i.e. threaded, within a CD ($\alpha$-CD may thread two monomers each, $\gamma$-CD may thread up to four. \textbf{(c-d)} same as \textbf{a-b} for PEG(Azo)$_2$. \textbf{(e)} Average fraction of PEG and CDs in the largest connected component (C.C) -- PEGs (left black axis, circle points), CD clusters (right red axis, square points) for $\gamma$-CD (orange) and $\alpha$-CD (green). \textbf{(f)} Degree of percolation of network, defined as the average number of connected components in the simulated network at steady state, as a function of different [CD] to [PEG] ratio.}
    \label{fig:5}
\end{figure*}

Having established that the CD crystalline domains dominate crosslinking and network formation in pPR gels, we questioned what contribution the double-threading of $\gamma$-CD has on the creation of stable gels. To test this, we considered $\alpha$-CD which can only accommodate a single PEG chain due to its smaller cavity~\cite{Harada1992,Harada1990}.

Bulk rheology performed on samples made with $\alpha$-CD broadly displayed the same behaviour as with $\gamma$-CD discussed above. However, we observed that the aggregation was slower (see SI, Fig.~S14) and that samples took longer to equilibrate. 
In Figure \ref{fig:5}a, we show that the system with PEG(OH)$_2$ and $\alpha$-CD displays a sol-gel transition at $>$4 CD per chain, and that all $\alpha$-CD pPRs are stiffer than the $\gamma$-CD pPRs. We explain this higher stiffness as follows: the crystalline aggregation of the small $\alpha$-CD acts to stabilise regions of CD that might act like crosslinks, adding stiffness. The crystallisation of $\alpha$-CD with PEG (with no aromatic end-group, i.e. -OH or -COOH) is also closer packed, it is known to pack via a hexagonal unit cell verses the cubic unit cell of $\gamma$-CD~\cite{Ohmura2004, Sakai2011}. The random coil of a sufficiently long polymer has been previously described as a kinetic speed-bump, preventing the translocation of $\alpha$-CD to the middle of the chain, and thus slowing bulk aggregation~\cite{Lin2021}. Therefore, we can explain the gel formation at 8:1 [CD]:[PEG] ratio of the PEG(OH)$_2$ system as a consequence of the slower translocation of $\alpha$-CD with respect to $\gamma$-CD~\cite{Takahashi2009}, which suppresses aggregation and in turn favours the formation of micro-crystalline domains that act as crosslinks. 

On the contrary, $\gamma$-CDs can more easily translocate on PEG chains due to their larger cavities. Ease of translocation, in addition to their additional threading site, acts to accelerate aggregation and precipitation. The second threading in $\gamma$-CD also provides a method to kinetically trap rings more easily, as a second threading event blocks de-threading whether parallel or branching. Aggregation with $\gamma$-CD is therefore strongly kinetically driven. In $\alpha$-CD pPRs the aggregation may only take place with the onset of crystallisation of regions of CD. The difference in rate of aggregation between CD size was clearly seen in turbidity measurements with PEG(OH)$_2$ in Fig.~S14.

The impact of ring size on stabilisation of regions of threaded CDs in PEG(OH)$_2$ pPRs is supported by Fig.~\ref{fig:5}b, as $^1$H NMR analysis shows intermediate ``found'' threaded ratios of $\alpha$-CD and not the cooperative increase in coverage exhibited by $\gamma$-CD.

Gel formation at lower feed ratios of CD with PEG(Azo)$_2$ is still seen with $\alpha$-CD, and a preference for a branching morphology remains prominent, as described in Figure \ref{fig:5}c. Simulations show that increasing $\alpha$-CD equivalents yields an increase in the fraction of PEGs in the largest connected component, and decreases in the number of CD clusters and number of connected components (Fig.\ref{fig:5}e-f). This suggests first an aggregation into clusters and then an increase in the number of clusters and size of clusters, as these clusters connect and approach percolation (Fig.S23).
Indeed, we show $\alpha$-CD networks featuring PEG(Azo)$_2$ follow the same gelation behaviour of the larger $\gamma$-CD homologue, with a shared gel-forming coverage at around 8 threaded sites (Fig.~\ref{fig:5}c-d).

These results help illustrate the promise of the ``figure zero'' PEG(\textbf{Azo})$_2$ system, as offers tunable stiffness with a better atom efficiency than $\alpha$-CD. The ``figure zero'' serves not as an exotic building block, building new network structures; instead, it serves to build the same types of structures as the $\alpha$-CD equivalent, as governed by the chain-end ring interaction, but at lower, and more controllable, stoichiometry, in turn potentially displaying larger tensile and stretching performance.  

\section{Conclusion}

\begin{figure*}[t]
    \centering
    \includegraphics[width=0.7\textwidth]{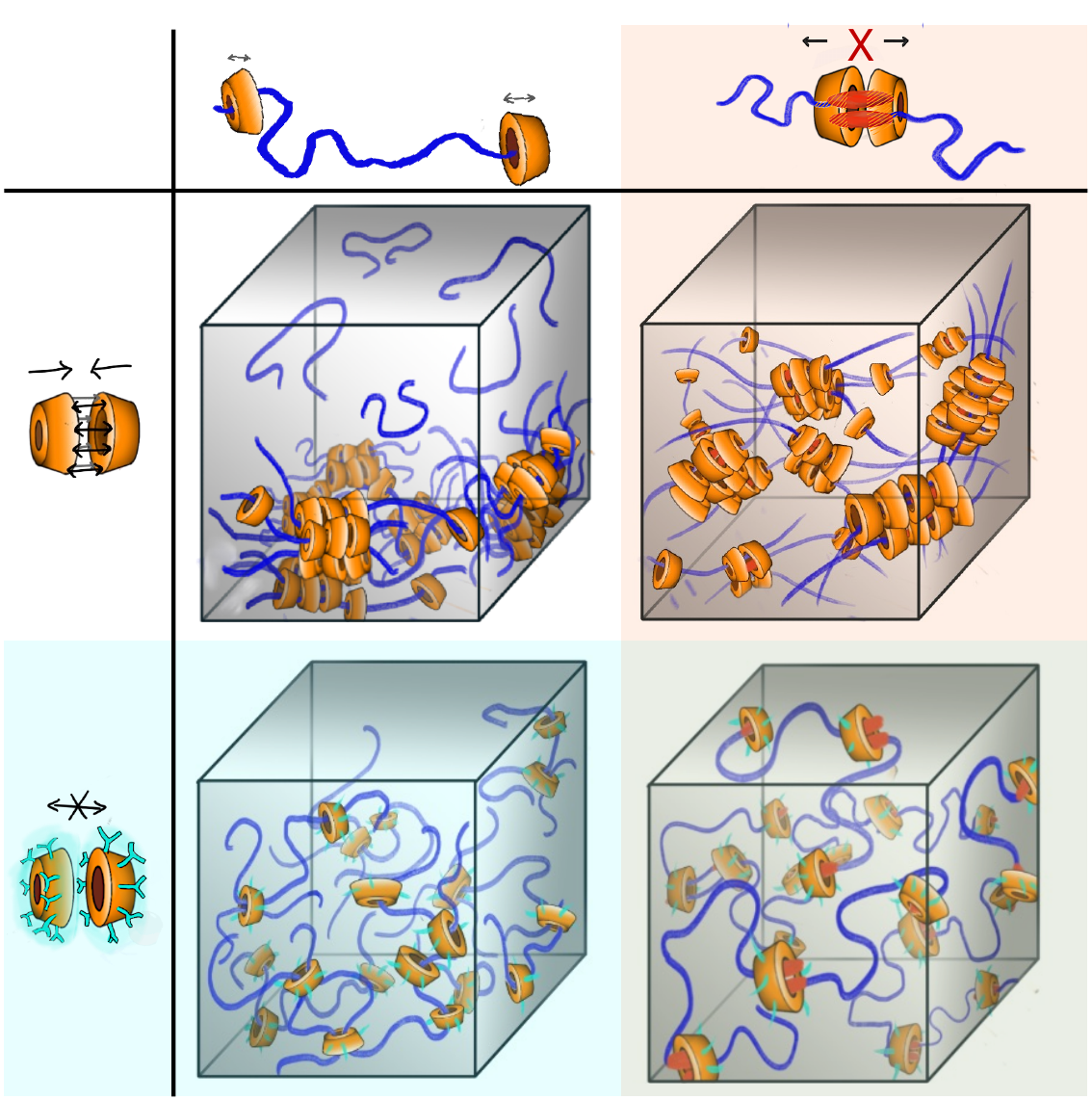}
    \caption{\textbf{Summary schematic representation of structures formed when intermolecular interactions are changed between pseudo-poly(rotaxane) components.} With strong ring-ring interaction ($\gamma$-CD, top row), and ``passivated'' ring-ring interactions (hp-$\gamma$-CD, bottom row); and free translocation from chain end (PEG(OH)$_2$, left column), and strong chain-end ring interaction (PEG(Azo)$_2$, right column).
    }
    \label{fig:6}
\end{figure*}

This work dissected the structure-function relationship of common ingredients in SRGs and should guide the rational development and design of novel SRGs.
We specifically focused on $\gamma$-CD as an emergent new component of SRGs~\cite{Tang2023}, which is relatively less studied compared with $\alpha$-CD. 

For each of these design variations, we employed bulk rheology to measure the macroscopic behaviour of pPRs and used molecular dynamics simulations to probe their microstructure, using other quantitative observables -- such as the number of CDs threaded by PEG chains -- to validate our simulated coarse-grained models against experiments.  

Several key findings emerge from this work. First, enhancement of the PEG-chain-end affinity by azobenzene functionalisation enables fine control over threading ratio and network stiffness, avoiding the abrupt, first-order-like transition seen with native PEG(OH)$_2$, Fig.~\ref{fig:3}. Such control allows gel formation at remarkably low stoichiometries, representing a promising strategy for maximising stretchability by preserving slidable length.  

Second, passivated $\gamma$-CD confirms that CD crystalline domains, rather than double-threading, are the principal contributor to crosslinking: hp-$\gamma$-CD show liquid-like rheology in both PEG(OH)$_2$ and PEG(Azo)$_2$ formulations (Fig.~\ref{fig:4}). We therefore propose that both $\alpha$- and $\gamma$-CD domains function as multivalent crosslinks, with their number governing the linear stress–strain response (Young's modulus) of any subsequent SRC gels. 

Finally, and counter-intuitively, $\gamma$-CD did not show evidence of yielding a significantly different gel-forming behaviour than $\alpha$-CD (Fig.~\ref{fig:5}). In other words, the rheological behaviour is dominated by the formation of CD domains, which in our pPRs appear insensitive (but for stoichiometry) on the size of CD  (Fig.~\ref{fig:5}). 

We would emphasise the value of PEG chain-end functionalisation for tunable, stiffer slide-ring materials. As shown in the work of Ke \textit{et al.}~\cite{Tang2023}, the stretching profile of SRGs built from these components was dissimilar to the typical J-curve seen in other slide-ring materials~\cite{Jiang2018, Wang2022}. The appeal of the ``figure zero'' motif, paired with a high binding end-group more broadly, arises from the resulting reinforcement through kinetically arresting the network. Chain-end functionalisation paired with the inclusion of a longer slidable region (via threaded PEG length) offers a preferable alternative to the softer (albeit potentially more extensible), less tunable structures that PEG(OH)$_2$ forms-- enabling a concurrent optimisation of stiffness and extensibility (see SI, Fig.~S25). A similar reinforcement, from strain-induced crystallisation, has been noted in SRGs -- where there is no copolymer -- however, the crystalline regions described in this work offer reinforcement to SRCs. Such a reinforcement offers functional, previously weak, hydrogels a route towards practical applications in products. 

There remains promising space for SRCs consisting of $\gamma$-CD; the ``figure-zero'' crosslinking motif also motivates the development of alternative high-binding groups, perhaps donor-acceptor charge-transfer complexes, for additional functionality or sensing. Understanding the threading process through this work is foundational for the rational design of such potentially complex systems. 

\section{Experimental Section}
\subsection{Materials}
All materials and consumables were purchased from commercial vendors and used without further purification. 

NMR were procured by Bruker Avance 400/500/600. All chemical shifts were calibrated in ppm via the residual proton peak in the relevant deuterated solvent, where CDCl3: $\delta$ = 7.26 ppm, DMSO-d6: $\delta$ = 2.50 ppm, D2O: $\delta$ = 4.79 ppm respectively. 

Time-dependent transmittance monitoring experiments were performed on a 
Hach DR3800
visible spectrophotometer equipped with a temperature controller using plastic cuvettes and calibrated before reading to the solvent and cuvette. 

Rheology was performed on a Kinexus ultra$+$ rheometer, equiped with either a 60mm smooth cone plate (samples $<$1Pa), and a 40mm sandblasted parallel plate with a gap height of 0.3mm (samples $>$1Pa).

\subsection{General procedure for the synthesis and rheological investigation of pPRs}
The PEG(Azo)$_2$ was prepared by an amended literature procedure (See Supplementary Material) \cite{Tang2023}. 
pPR synthesis was performed 
by the mixing of aqueous solutions of CD and PEG. 
Separate aqueous mother solutions of both PEG (26mM) and CD (258mM) were perpared. Aliquots of both solutions were loaded into luer-lock syringes and with the CD solution being diluted to the relevant concentration. Bubbles were then carefully removed from the syringe, the two were locked together with a luer-lock connector and mixed thoroughly between them by 20 full back-and-forth cycles. The pPR pre-gel solution was then (where viscosity allows) transferred into an Eppendorf and centrifuged to remove bubbles, after which the pre-gel solution was subsequently poured onto a rheometer's 
geometry. The 
solution was allowed to equilibrate on the plate and the time evolution of the fluid was tracked via time-dependent single-frequency sweeps (at 0.2Hz, and 0.5\%) within the linear regime (see SI-Rheological Investigations of pPR gels for details). The stiffness of the pPR was then determined by frequency sweeps at a constant oscillatory amplitude (0.5\%), again within the linear regime, to establish a plateau modulus, where applicable.
Amplitude sweeps were performed both to confirm the suitability of the amplitude at which the frequency sweeps were performed and collected to establish yield stress by sweeping the larger 0.1-100\% window at 1Hz.

\section*{Supporting Information}
Supporting Information is available from the authors upon request.

\section*{Author Contributions}
Author contributions to this work are as follows: 
E.S. contributed to the conceptualisation, investigation, methodology, formal analysis, writing – original draft, and writing – review \& editing. 
A.B. contributed to investigation, software, methodology, formal analysis, and writing – review \& editing.
S.W. contributed to the formal analysis and methodology.
N.Z. provided funding.
P.L. provided resources. 
C.K. contributed to the conceptualisation, methodology, resources, supervision, and writing – review \& editing.
D.M. contributed to the conceptualisation, supervision, resources, writing – original draft, and writing – review \& editing.

\begin{acknowledgments}
DM acknowledges the Royal Society and the European Research Council (grant agreement No 947918, TAP) for funding.
ES acknowledges the Engineering and Physical Research Council (EPSRC),  Soft Matter for Formulation and Industrial Innovation centre for Doctoral Training (SOFI$^2$ CDT), and Reckitt Benckiser Group PLC for funding (EP/S023631/1).
Authors would also like to acknowledge the groups of PL, DM, SOFI CDT cohort members and F. Schaufelberger for scientific discussion and feedback.
\end{acknowledgments}

\bibliographystyle{apsrev4-2}
\bibliography{library}

\end{document}